\documentclass{elsart3-1}

\usepackage{amssymb,amsmath,epsfig,color}
\usepackage[english,francais]{babel}
\usepackage{epstopdf}
\usepackage{framed}

\newcommand{\remfigure}[1]{}

\newcommand{\bu}{\boldsymbol{u}}

\newcommand{\bx}{\boldsymbol{x}}
\newcommand{\bp}{\boldsymbol{p}}
\newcommand{\bq}{\boldsymbol{q}}

\newcommand{\scrap}[1]{}

\def\contract{\makebox[1.2em][c]{\mbox{\rule{.6em}
{.01truein}\rule{.01truein}{.6em}}}}
\newcommand{\bfi}{\bfseries\itshape}

\newtheorem{theorem}{Theorem}[section]

\newtheorem{e-proposition}[theorem]{Proposition}
\newtheorem{corollary}[theorem]{Corollary}
\newtheorem{e-definition}[theorem]{Definition\rm}

\renewcommand{\theequation}{\arabic{equation}}
\def\og{\leavevmode\raise.3ex\hbox{$\scriptscriptstyle\langle\!\langle$~}}
\def\fg{\leavevmode\raise.3ex\hbox{~$\!\scriptscriptstyle\,\rangle\!\rangle$}}

\journal{the Acad\'emie des sciences}
\begin{document}
\centerline{}
\begin{frontmatter}


\selectlanguage{english}
\title{Geometric dissipation in kinetic equations}

\bigskip


\selectlanguage{english}
\author[authorlabel1]{Darryl D. Holm},
\ead{d.holm@imperial.ac.uk}
\author[authorlabel2]{Vakhtang Putkaradze},
\ead{putkarad@math.colostate.edu}
\author[authorlabel3]{Cesare Tronci}
\ead{cesare.tronci@imperial.ac.uk}

\address[authorlabel1]{Department of Mathematics, Imperial College London, London
SW7 2AZ, UK\\
Computer and Computational Science Division,
Los Alamos National Laboratory, Los Alamos, NM, 87545 USA}
\address[authorlabel2]{Department of Mathematics, Colorado State University,
Fort Collins, CO 80523 USA\\
Institute for Theoretical Physics,
 Universit\"at K\"oln, Zuplicher Str. 77, 50968 K\"oln,
Germany}
\address[authorlabel3]{Department of Mathematics, Imperial College London, London
SW7 2AZ, UK\\
TERA Foundation for Oncological Hadrontherapy, 
11 V. Puccini, Novara 28100, Italy
}


\bigskip
\scrap{
\begin{center}
{\small Received *****; accepted after revision +++++\\
Presented by £££££}
\end{center}
}

\begin{abstract}
\selectlanguage{english}
A new symplectic variational approach is developed for modeling dissipation in kinetic equations. This approach yields a double bracket structure in phase space which generates kinetic equations representing coadjoint motion under canonical transformations. The Vlasov example admits measure-valued single-particle solutions. Such solutions are reversible. The total entropy is a Casimir, and thus it is preserved.\\ 
{\it To cite this article: D.D.Holm, V. Putkaradze and C. Tronci, 
C. R. Acad. Sci. Paris, Ser. I XXX (2007).}

\vskip 0.5\baselineskip

\selectlanguage{francais}
\noindent{\bf R\'esum\'e} \vskip 0.5\baselineskip \noindent
Une nouvelle approche est propos\'ee pour modeliser la dissipation dans les \'equations cin\'etiques. Cette approche produit une structure \`a double crochet dans l'espace des phases qui conduit aux 
\'equations cin\'etiques d'une dynamique coadjointe apr\`es transformations canoniques. L'exemple de Vlasov admet alors des solutions pour particule unique. Ces solutions sont r\'eversibles; l'entropie totale est un Casimir et est donc pr\'eserv\'ee.

{\it Pour citer cet article~:D.D.Holm, V. Putkaradze and C. Tronci, 
C. R. Acad. Sci. Paris, Ser. I XXX (2007). }

\end{abstract}
\end{frontmatter}

\selectlanguage{francais}
\section*{Version fran\c{c}aise abr\'eg\'ee}
Une nouvelle approche est propos\'ee pour la mod\'elisation des ph\'enom\`enes dissipatifs dans les \'equations cin\'etiques \cite{Fokker-Plank,Vl1961}. Cette construction est r\'ealis\'ee de telle sorte que la g\'eom\'etrie de la variable dynamique est pr\'eserv\'ee: en particulier, nous consid\'erons l'\'equation de Vlasov \cite{Vl1961} comme mod\`ele naturel de conservation dans l'espace des phases. De plus 
nous introduisons une quantit\'e particuli\`ere nomm\'ee ``mobilit\'e'', inspir\'ee par analogie avec la loi de Darcy pour syst\`emes continus avec auto-agr\'egation
\cite{HP2005,HP2006,HP2007}. Dans ce cas, on introduit la mobilit\'e comme le facteur de proportionnalit\'e entre la force agissant sur les particules et leur vitesse. Nous nous
int\'eressons ainsi \`a une forme de dissipation qui peut g\'en\'eraliser les ph\'enom\`enes d'auto-agr\'egation aux syst\`emes cin\'etiques dans l'espace des phases.
 
Une telle approche produit une structure \`a double crochet \cite{BlKrMaRa1996}  dans l'espace des phases similaire \`a celle pr\'esent\'ee dans la litt\'erature pour la mod\'elisation de certains syst\`emes astrophysiques \cite{Ka1991}. Cette structure g\'en\`ere une dynamique coadjointe r\'eversible (\'Eq. \ref{Vlasov-diss}) {\it via} l'action des transformations canoniques. On trouve finalement que toutes les fonctionnelles de la distribution sont de type Casimir et que l'entropie est pr\'eserv\'ee
(Proposition \ref{Casimir-conserv}). 

L'innovation de notre approche se voit dans le r\^ole que la mobilit\'e peut jouer comme op\'eration de filtre (o moyenne) sur la fonction de distribution des particules. Par cons\'equent on d\'efinit la mobilit\'e comme une fonctionnelle de la distribution des particules. Ce fait conduit \`a l'existence de la solution de particule unique  qui n'est pas pr\'esente dans les anciennes approches et repr\'esente le r\'esultat principal de cet article (Th\'eor\`eme \ref{singparticles}).


\selectlanguage{english}
\section{Introduction}
\label{}
Non-linear dissipation in physical systems can modeled by the sequential application of  two Poisson brackets, just as in magnetization dynamics \cite{Gilbert}. A similar double bracket operation for modeling dissipation has been proposed for the Vlasov equation. Namely,
\begin{equation}
\frac{\partial f}{\partial t}+ \left[\,f\,,\,\frac{\delta H}{\delta f}\right]
=
\alpha \left[\,f\,,\,\left[\,f\,,\,\frac{\delta H}{\delta f}\,\right] \right]
\,,
\label{Kandrup-dbrkt}
\end{equation}
where $\alpha>0$ is a positive constant, $H$ is the Vlasov Hamiltonian  and 
$[ \cdot \, , \, \cdot ]$ is the canonical Poisson bracket. When $\alpha\to0$, this equation reduces the Vlasov equation for collisionless plasmas. For $\alpha>0$, this is the {\bfi double bracket dissipation} approach for the Vlasov-Poisson equation introduced in Kandrup \cite{Ka1991} and developed in Bloch {\it et al.} \cite{BlKrMaRa1996}.
This double bracket approach for introducing dissipation into the Vlasov equation differs from the standard Fokker-Planck linear diffusive approach \cite{Fokker-Plank}, which adds dissipation on the right hand side as the Laplace operator in the momentum coordinate $\Delta_p f$. 

 An interesting feature of the double bracket approach is that the resulting symmetric bracket gives rise to a metric tensor and an associated Riemannian (rather than symplectic) geometry for the solutions.  The variational approach also preserves the {\bfi advective} nature of the evolution of Vlasov phase space density, by coadjoint motion under the action of the canonical transformations on phase space densities.

 As Otto \cite{Ot2001} explained, the geometry
of disspation may be understood as emerging from a variational  principle. Here, we apply the variational approach to derive the following generalization of the double bracket structure in equation (\ref{Kandrup-dbrkt}) that recovers previous cases for particular
choices of modeling quantities, 
\begin{equation}
\frac{\partial f}{\partial t}
+
\left[\,f\,,\,\frac{\delta H}{\delta f}\,\right]
\,=\,
\left[\,f\,,\,\left[\,\mu(f)\,,\,\frac{\delta E}{\delta f}\,\right]
\,\right]
\,.
\label{Vlasov-diss}
\end{equation}
Eq. (\ref{Vlasov-diss}) extends the double bracket operation in (\ref{Kandrup-dbrkt}) and reduces to it when $H$ is identical to $E$ and $\mu(f)=\alpha\,f$.
The form (\ref{Vlasov-diss}) of the Vlasov equation with dissipation allows for more general mobilities than those in \cite{BlKrMaRa1996,Ka1991,Ka1984,Mo1984}.
For example, one may choose $\mu[f]=K*f$ (in which $*$ denotes convolution in phase space). As in \cite{HoPuTr2007} the smoothing operation in the definition of $\mu(f)$ introduces a fundamental length scale (the filter width) into the dissipation mechanism.  Smoothing has the added advantage of endowing (\ref{Vlasov-diss}) with the one-particle solution as its singular solution. 
The generalization Eq. (\ref{Vlasov-diss}) may also be justified by using thermodynamic and geometric arguments \cite{HoPuTr2007}. 
In particular, this generalization extends the classic Darcy's law (velocity being proportional to force) to allow the corresponding modeling at the microscopic statistical  level.

\section{Dissipation for kinetic equations}



We aim to model dissipation in Vlasov kinetic systems through a suitable generalization of Darcy's law. 
Indeed, we believe that the
basic ideas of Darcy's Law in configuration space can be transferred to a phase space treatment giving rise to the kinetic description of self-organizing
collisionless multiparticle systems. 
\scrap{
The main issue here  is to accurately
consider  not only the geometry of particle distribution,
but also the structure of the phase space itself. As is well known, properties
of the phase space (momentum and position) are completely different from the configuration space (position only)
because of the symplectic relation between the momentum and position. This
structure of the phase space warrants a suitable modification of the diamond
operator.
}
  In what follows, we
will construct kinetic equations for geometric order parameters that respect the symplectic nature of the phase space by considering the Lie algebra of
generating functions of canonical transformations (symplectomorphisms).

The first step is to establish how a geometric quantity evolves in phase space, so that the symplectic nature of its evolution is preserved.  For this,  we regard the action of the symplectic algebra as an action of the generating functions $h$ of $\kappa$, rather than vector fields. The action is formally expressed as $\kappa\, h=\pounds_{X_h}\,\kappa$.  The dual operation of the action which we 
denote by $\star$ is then defined as $\langle \,\kappa\star \zeta,\, h \,\rangle = \langle\, \kappa,-\, \mathcal{L}_{{X_{\!h}}}\,\zeta \,\rangle$. Here $X_h(q,p)$ is the Hamiltonian
vector field  generated by a Hamiltonian function $h(q,p)$ through
the definition $X_h \contract \,\omega:=dh$. Notice that the star operation takes values in the space $\mathcal{F}^*$ of phase space densities  $\kappa\star \zeta\in\mathcal{F}^*$. In the particular case of interest here, $\kappa$ is the phase space density $\kappa=f\, dq\wedge dp$ and $\zeta=g$,  a function on phase space. In this case, the star operation is simply minus the canonical Poisson bracket, $\kappa\star g =[f,g]\,dq\wedge dp$.

We shall first employ these considerations to 
find the \emph{purely dissipative} part of the kinetic equation for a particle
density on
phase space. We choose variations of the form $\delta f=-\pounds_{{\,X_{h}(\phi)}} \,\,\mu(f)=-\,[\mu(f),h(\phi)]$ with 
$h(\phi)=(\phi \star f)^\sharp =[\phi \, , \, f]$ where $(\,\cdot\,)^\sharp$ in $(f \star \phi)^\sharp$ transforms a phase space density into a scalar function. The operation $(\,\cdot\,)^\sharp$ will be understood in the pairing below. We then follow the steps:
\begin{eqnarray*}\hspace{-8.5mm}
\left\langle \phi,\frac{\partial f}{\partial t} \right\rangle
= 
\left\langle \frac{\delta E}{\delta f},\delta f \right\rangle
=
 \left\langle \frac{\delta E}{\delta f}, -\Big[\mu(f),h(\phi)\Big] \right\rangle
=
 \Bigg\langle \left[\mu(f) , \frac{\delta E}{\delta f}\right], \bigg[\phi ,
f \bigg]
\Bigg\rangle
=\Bigg\langle \phi , \left[f,\left[\mu(f),\frac{\delta E}{\delta f}\right]\right] \Bigg\rangle .
\end{eqnarray*}
Therefore, a functional $F(f)$ satisfies the following evolution equation in bracket notation,
\begin{eqnarray}
\frac{d F}{dt}
=
\left\langle \frac{\partial f}{\partial t} \,,\, 
\frac{\delta F}{\delta f} \right\rangle
=
-\,\Bigg\langle \left[\,\mu(f)\,, \frac{\delta E}{\delta f}\right],\, \left[\,f\,,\frac{\delta F}{\delta f}\right] \Bigg\rangle
=:
\{\!\{\,E\,,\,F\,\}\!\} 
\,.
\label{diss-bracket}
\end{eqnarray}

The mobility $\mu$ and dissipation energy functional $E$ appearing in (\ref{diss-bracket})
are modeling choices and must be selected based on the additional input from physics. The bracket (\ref{diss-bracket}) reduces to
Kandrup's dissipative bracket for the modeling choice of $\mu(f)=\alpha f$ with some $\alpha>0$ \cite{Ka1991}. The dissipation energy $E$ in Kandrup's paper was taken to be the Vlasov Hamiltonian (see below), but in our approach it also may be taken as a modeling choice. This extra freedom allows for more flexible interpretation and treatment of the dissipation process. 

\begin{e-proposition}\label{energy-dissip}
There exist choices of mobility $\mu[f]$ for which the bracket (\ref{diss-bracket}) dissipates energy $E$.
\end{e-proposition}
\noindent
{\bf Proof.} The dissipative bracket in equation (\ref{diss-bracket}) yields $\dot{E}=\{\!\{\,E\,,\,E\,\}\!\}$ which is negative when $\mu[f]$ is chosen appropriately.  For example, $\mu[f]=f M[f]$, where $M[f] \geq 0$ is a non-negative scalar functional of $f$. (That is, $M[f]$ is a number.) 

\begin{rem}
The dissipative bracket (\ref{diss-bracket}) satisfies the Leibnitz rule for the derivative of a product of functionals. In addition, it allows one to reformulate the equation (\ref{Vlasov-diss}) in terms of flow on a Riemannian manifold with a metric defined through the dissipation bracket, as discussed in more detail in \cite{HP2007}.
\end{rem}\smallskip

\begin{e-proposition}[Casimir functionals]
\label{Casimir-conserv}
For an arbitrary smooth function $\Phi$ the functional $C_\Phi=\int\! \Phi(f)\, dq\wedge dp$ is preserved for any energy functional $E$. 
\end{e-proposition}
\noindent
{\bf Proof.}
It suffices to calculate the bracket 
\begin{equation}
\frac{dC_\Phi}{dt}=\{\{C_\Phi,E\}\}
:=-\,\Bigg\langle 
\left[\,\mu(f)\,, \frac{\delta E}{\delta f}\right],\, \left[\,f\,,\frac{\delta C_\Phi}{\delta f}\right] \Bigg\rangle\\
=
-\, \Bigg\langle \left[\,\mu(f)\,, \frac{\delta E}{\delta f}\right],\, \bigg[\,f\,,\,\Phi'(f)\,\bigg] \Bigg\rangle=0
. 
\end{equation}
\begin{corollary}\label{entropy-conserv}
The entropy functional $S=\int\!f\,\log f$ is preserved for any energy functional $E$. 
\end{corollary}
\begin{rem}
The existence of Casimirs and the corresponding preservation of any entropy defined solely in terms of $f$ arises because the dissipative bracket (\ref{diss-bracket}) generates coadjoint motion, which is  {\bfi reversible}. This property is shared with Kandrup's bracket, which is recovered for $\mu(f)=\alpha f$ for constant $\alpha>0$.
\end{rem}

\section{Dissipation for kinetic moments: the Kupershmidt-Manin bracket and
Darcy's law}
In this section we show how Eq. (\ref{Vlasov-diss}) leads very naturally to a nonlocal form of Darcy's law. In order to show how this equation is recovered, we introduce the Lie-Poisson structure for kinetic moments, also known as Kupershmidt-Manin structure \cite{KuMa}.
We proceed by considering a one-dimensional configuration space; an extension to higher dimensions would also be possible by considering the treatment in \cite{GiHoTr05}.

\smallskip
As a general result \cite{Gi,GiHoTr05,GiHoTr07}, the equations for the moments of the Vlasov equation are a Lie-Poisson system. The $n$-th moment is defined as
\[
A_n(q):=\int p^n\, f(q,p)\, dp\,.
\]
and the dynamics of these quantities is regulated by the {Kupershmidt-Manin structure}
\[
\{F,G\}=
\left\langle A_{m+n-1},\,
\left[\!\!\left[\frac{\delta F}{\delta A_n},\frac{\delta G}{\delta A_m}\right]\!\!\right]
\right\rangle
\,,
\]
where summation over repeated indices is omitted and the Lie bracket $\left[\!\left[\cdot,\cdot\right]\!\right]$ is defined as
\[
\left[\!\left[\alpha_m,\,\beta_n\right]\!\right]\,=\,
n\,\beta_n(q)\,\alpha_m'(q)
-
m\,\alpha_m(q)\,\beta_n^{\,\prime}(q)
\,=:\,
\textsf{\large ad}_{\alpha_m}\, \beta_n
\]
The moment equations are
\[
\dot{A}_n=-\,\textsf{\large ad}^*_{\beta_n}\,A_{m+n-1}=
-\left(  n+m\right) \, A_{n+m-1}\,\frac{\partial
\beta_{n}}{\partial q}
-
n\,\beta_{n}\,\frac{\partial
A_{n+m-1}}{\partial q}
\,,
\]
where the ${\sf ad}^*$ operator is defined by $\langle\, {\sf ad}^*_{\beta_n} \,A_k,\,\alpha_{k-n+1}\,\rangle:=
\langle\, A_k,\,{\sf ad}_{\beta_n}\,\alpha_{k-n+1}\,\rangle$.
\smallskip

At this point one can consider the following Lie algebra action on Vlasov densities
\[
\beta_n\,f:=\text{\it\large\pounds}_{X_{p^n\beta_n}}f=\big[\,f,\,p^n\beta_n\big]
\qquad\text{ (no sum)}
\]
which is obviously given by the action of the Hamiltonian function $
h(q,p)=p^n\beta_n(q)$.
Now, the dual action is given by
\begin{align*}
\Big\langle f\,\text{\large$\star$}_{n}\, g,\,\beta_n\Big\rangle:=
\Big\langle f,\, \beta_n\, g\Big\rangle &=
\Big\langle f\!\star g\,,\, p^n \beta_n(q) \Big\rangle =
 \left\langle \int\{f, g\}\,p^n\,dp\,,\,\beta_n \right\rangle
\end{align*}
and the dissipative bracket for the moments (\ref{diss-bracket}) is written in this notation as
\begin{align*}
\{\!\{\,E\,,\,F\,\}\!\}
&= 
-\,\Bigg\langle \int\! p^n\left[\,\mu[f]\,, \frac{\delta E}{\delta f}\right]dp,\,\int\!
p^n \left[f\,,\frac{\delta F}{\delta f}\right]dp \Bigg\rangle
\\&=
-\left\langle\textsf{\large ad}^*_{\beta_k}\, \widetilde{\mu}_{\,k+n-1},\,
\left(\textsf{\large ad}^*_{\alpha_m}A_{m+n-1}\right)^{\sharp}\,\right\rangle
\end{align*}
where we have substituted $\delta E/\delta f=p^k\beta_k$ and $\delta F/\delta f=p^m\alpha_m$ and $\widetilde{\mu}_s(q):=\int\! p^s \mu[f]\,dp$. 

Thus the purely dissipative moment equations are
\[
\dot{A}_n=\textsf{\large ad}^*_{\gamma_m}A_{m+n-1}
\qquad\text{with}\qquad
\gamma_m:=\left(\textsf{\large ad}^*_{\beta_k}\, \widetilde{\mu}_{\,k+m-1}\right)^\sharp
\]
If we now write the equation for $\rho:=A_0$ and
consider only $\gamma_0$ and $\gamma_1$, we recover the following form of Darcy's law
\[
\dot\rho=\,\textsf{\large ad}^*_{\gamma_1}\rho=\,
\frac{\partial}{\partial q}\!\left(\rho\,\mu[\rho]\,\frac{\partial}{\partial q}\frac{\delta E}{\delta
\rho}\right)
\]
where we have chosen $E=E[\rho]$ and $\widetilde{\mu}_0=\mu[\rho]$, so that $\,\gamma_1=\widetilde{\mu}_0 \,\partial_q\beta_0$.

\subsection{Special cases}
Two interesting cases may be considered at this point. In the first case one makes Kandrup's choice in (\ref{Kandrup-dbrkt}) for the mobility at the kinetic level $\mu[f]=f$, so that Darcy's law is written as
\[
\dot\rho=\frac{\partial}{\partial q}\!\left(\rho^2\,\frac{\partial}{\partial q}\frac{\delta E}{\delta
\rho}\right)\,.
\]
Kandrup's case applies to the dissipatively induced instability of galactic dynamics \cite{Ka1991}. The previous equation is the Darcy law description of this type of instability. 
In the second case, one considers the mobility $\mu[\rho]$ as a functional of $\rho$ (a number), leading to the equation
\[
\dot\rho=\mu\,\frac{\partial}{\partial q}\!\left(\rho\,\frac{\partial}{\partial q}\frac{\delta E}{\delta
\rho}\right)\,,
\]
which leads to the classic energy dissipation equation, $dE/dt=-\,\langle\mu\rho|\frac{\partial}{\partial q}\frac{\delta E}{\delta
\rho}|^2 \rangle $.

\subsection{Summary}
This section has provided a consistent derivation of Darcy's law from first principles in kinetic theory, obtained by inserting dissipative terms into the Vlasov equation which respect the geometric nature of the system. This form of the Darcy's law has been studied and analyzed
in \cite{HP2005,HP2006}, where it has been shown to possess emergent singular solutions ({\it clumpons}), which form spontaneously and collapse together in a finite time, from any smooth confined initial condition.

\section{A dissipative Vlasov equation}

The discussion from the previous sections produces an interesting
opportunity for the addition of dissipation to kinetic equations. This opportunity arises from noticing that the dissipative bracket derived here could just as well be used with any type of evolution
operator. 
In particular, we may consider introducing  our bracket to modify Hamiltonian dynamics as in the approach by Kaufman \cite{Ka1984} and Morrisson \cite{Mo1984}.
 In particular, the dissipated energy may
naturally be associated with the Hamiltonian arising from the corresponding Lie-Poisson theory for the evolution of a particle distribution function
$f$. Therefore, we write the total dynamics generated by any functional 
$F(f)$ as $\dot{F}=\left\{F,H\right\}+\left\{\left\{F,E\right\}\right\}$ where  
$\left\{\cdot \, , \, \cdot \right\}$ represents the Hamiltonian part of the dynamics. 
This gives the {\it dissipative Vlasov equation} of the form (\ref{Vlasov-diss}) with $E=H$, where $H(f)$ is the Vlasov Hamiltonian. We illustrate these ideas by computing the singular (measure-valued) solution of equation (\ref{Vlasov-diss}), which represents the reversible motion of a single particle.

\begin{theorem}\label{singparticles}
Taking $\mu(f)$ to be an arbitrary function of the smoothed distribution
$\bar{f}=K*f$ for some kernel $K$  allows for single particle solutions $f=\sum_{i=1}^Nw_i
\delta(q - {Q}_i(t) ) \delta(p- {P}_i(t))$. 
The single particle dynamics is governed by canonical equations with Hamiltonian given by
\[
\mathcal{H}=\left(\frac{\delta H}{\delta f}-
\left[\mu\left(f\right),\frac{\delta H}{\delta f}\right]\right)_{(q,p)=(Q_i(t),P_i(t))}
\]
\end{theorem}
\noindent
{\bf Proof.}
Let us write the equation of motion (\ref{Vlasov-diss}) in the following compact form
\[
\frac{\partial f}{\partial t}=-\,\left[\,f,\,\mathcal{H}\,\right]
\,,\qquad\text{ with }\quad
\mathcal{H}
:=
\left(\frac{\delta H}{\delta f}-
\left[\mu\left(f\right),\frac{\delta H}{\delta f}\right]\right)
\]
and substitute the single particle solution ansatz
$
f(q,p,t)\,=\,\sum_i w_i\,\delta(q-Q_i(t))\,\delta(p-P_i(t))
$.
Now take the pairing with a phase space function $\phi$ and write
$
\langle\, \phi,\,\dot{f}\,\rangle=-\left\langle\, 
\left[\,\phi,\,\mathcal{H}\,\right],\,
f\,\right\rangle
$.
Evaluating on the delta functions proves the theorem.
\begin{rem}
The quantity $-
[\mu\left(f\right),{\delta H}/{\delta f}]$
plays the role of a Hamiltonian for the advective dissipation process by coadjoint motion. This Hamiltonian is constructed from the momentum map $J$
defined by the $\star$ operation (Poisson bracket). That is, $J_h(f,g) 
= \langle g, -\pounds_{X_h}f\rangle 
= \langle g, [h,f]\rangle 
= \langle h, [f,g]\rangle 
= \langle h, f \star g\rangle$.
\end{rem}
\scrap{
\begin{equation*}
\dot{w}_i=0
\,,\quad
\dot{Q}_i=\frac{\partial\mathcal{H}}{\partial P_i}
\,,\quad
\dot{P}_i=
-\,\frac{\partial\mathcal{H}}{\partial Q_i}
\end{equation*}
where one evaluates $\mathcal{H}$ at the phase point $(q,p)=(Q_i(t),P_i(t))$.
}

\section{Discussion and Conclusions}
\scrap{
In this paper, we have developed a geometric approach to the derivation of
the kinetic equations and considered as an example a dissipative Vlasov
equation.
We have achieved this result through geometric generalization of the Darcy's law  for the symplectic $(\bq,\bp)$ phase space. Continuing our geometric
approach, it is possible to derive a dissipative Vlasov equation for the
particles with internal degrees of freedom, for example, energy $E$  dependent
on the mutual position and orientation of particles in the three dimensional
space.  
}

This paper has developed a new symplectic variational approach for modeling dissipation in kinetic equations based upon a double bracket structure in phase space. We focused our approach on the Vlasov example and found that the Vlasov case allows single-particle solutions, provided the mobility in the dissipation is a functional of the phase space distribution function. Moreover, we have shown how this approach recovers a nonlocal form of Darcy's law by using the Kupershmidt-Manin structure for kinetic moments. In general, it is also possible to extend our theory to the evolution of an arbitrary geometric quantity defined on any  smooth manifold \cite{HoPuTr2007}. For example, the restriction of the geometric formalism for symplectic motion considered here to cotangent lifts of diffeomorphisms recovers the corresponding results for fluid momentum. One may also extend the present phase space treatment to include an additional set of dimensions corresponding to statistical internal degrees of freedom (order parameters, or orientation dependence) carried by the microscopic particles, rather than requiring them to be point particles. This is a standard approach in condensed matter theory, for example in liquid crystals, see, e.g., \cite{Ch1992,deGePr1993}.
\smallskip

\begin{rem}
Being a special case of the dissipative bracket (\ref{diss-bracket}) presented in this paper, Kandrup's double bracket in (\ref{Kandrup-dbrkt}) also possesses the Casimirs found in Proposition \ref{Casimir-conserv}. However, the evolution under Kandrup's  double bracket does not allow single particle solutions. 
\end{rem}
\smallskip

\begin{rem}
Had we chosen variations of the form $\delta f=-\pounds_{{\,X_{h}(\phi)}} \,\,f=-\,[f,h(\phi)]$ with $h(\phi)=\mu(f) \star \phi$ $=\![\mu(f) \, , \, \phi]$ and followed the same steps as those in deriving (\ref{diss-bracket}), we would have obtained a different  dissipative double bracket. It would have the same form as (\ref{diss-bracket}), but with $\mu(f)\leftrightarrow f$ switched in the corresponding entries. These two choices have different thermodynamic implications. In particular, the calculation in the proof of Proposition \ref{entropy-conserv} would give entropy dynamics of the form
\begin{equation*}
\frac{dS}{dt}=\{\{S,E\}\}
=
-\, \Bigg\langle \left[\,f\,, \frac{\delta E}{\delta f}\right],\, \bigg[\,\mu(f)\,,\,\log f\,\bigg] \Bigg\rangle
=
-\, \Bigg\langle \frac{\mu(f)}{f}
\,,
\Bigg[\,f\,, \left[\,f\,, \frac{\delta E}{\delta f}\right] \Bigg] \Bigg\rangle
\ne0
. 
\end{equation*}
For entropy increase, this alternative variational approach would require $\mu(f)$ and $E(f)$ to satisfy an additional  condition (e.g., $\mu(f)/f$ and $\delta E/\delta f$ functionally related). However, the Vlasov dissipation induced in this case would not allow the reversible single-particle solutions, because of the loss of information associated with entropy increase. 
\end{rem}




\section*{Acknowledgements}
DDH and VP were partially supported by NSF grant NSF-DMS-05377891.
DDH was also partially supported  by the US Department of Energy,
Office of Science, Applied Mathematical Research and the Royal Society Wolfson Research Merit Award. VP is grateful for
the support of the Humboldt foundation and the hospitality of the
Institute for Theoretical Physics, University of Cologne where this
project was completed. We would also like  to thank the European Science Foundation for partial support through the MISGAM program.
Finally, we thank C. Josserand for helpful discussions. 
\bigskip


\remfigure{
\newpage

\noindent
{\bf List of modifications, following referee's suggestions}

\begin{itemize}
\item
Editorial, e.g., ``Darcy's law'' throughout and typos removed.
\item
Subsection 3.1 to give explicitely some specific forms of PDEs generated by this machinery.
\item
Subsection 3.2 to give an intermediate summary.
\end{itemize}
}

\end{document}